# ATLAS IBL: a challenging first step for ATLAS Upgrade at the sLHC

**Alessandro La Rosa[1] on behalf of the ATLAS Collaboration**
*DPNC, University of Geneva*
*CH 1211 Geneva 4, Switzerland*
*E-mail:* `alesandro.larosa@cern.ch`

With the LHC collecting data at 7 TeV, plans are already advancing for a series of upgrades leading eventually to about five times the LHC design luminosity some 10 years from now in the High Luminosity LHC (HL-LHC) project.

The upgrades for ATLAS detector will be staged in preparation for HL-LHC. The first upgrade for the Pixel Detector will be the construction of a new pixel layer, which will be installed during the first shutdown of the LHC machine foreseen in 2013-14.

The new detector, called the Insertable B-Layer (IBL) will be installed between the existing pixel detector and a new, smaller radius beam-pipe at the radius of 3.2 cm.

The IBL will require the development of several new technologies to cope with increased radiation and pixel occupancy and also to improve the physics performance through reduction of the pixel size and more stringent material budget. Two different and promising Silicon sensor technologies (planar n-in-n and 3D) are currently under investigation for the IBL.

An overview of the IBL module design and the qualification for these sensor technologies are presented in this proceeding. This proceeding also summarizes the improvements expected to the ATLAS detector at the HL-LHC.

---

[1] Speaker

## 1. Introduction

The ATLAS detector [1] is a general purpose detector at the Large Hadron Collider at CERN which was designed to be sensitive to a wide range of physics signatures to fully exploit the physics potential of the LHC collider at a nominal luminosity of $10^{34}$ cm$^{-2}$s$^{-1}$. After the successful commissioning and operation, it is planned to extend the LHC physics program by increasing about five times the luminosity in the so-called High-Luminosity LHC (HL-LHC) with a target of an integrated luminosity of 3000 fb$^{-1}$. The upgrades for the ATLAS Inner Detector (ID), which consists of a Pixel Detector [2], a Silicon micro-strip tracker (SCT) [3] and a gas-based transition radiation tracker (TRT) [4], are staged in view of the HL-LHC.

The first upgrade step (foreseen for the first long shutdown of LHC machine, 2013-14) consists in the construction of a new innermost Pixel Detector layer so-called Insertable B-Layer (IBL) [5] that will be installed together with a new beam pipe to maintain an excellent vertex detector performance and compensate possible inefficiencies of the current Pixel Detector.

Considering that the currently ID sub-detector would become inefficient during the expected HL-LHC operation scenario a replacement of the all ID is foreseen for the third long shutdown of LHC machine (~2022). The future ATLAS tracker thought for that replacement should consist in an all silicon-based system with new detector technologies. Its design is under investigation and the baseline consists in having the innermost region covered by Pixel Detector and the intermediate and outer radii region by micro-strip detectors. The region of the future Pixel Detector is subdivided into two domains: *outer pixel layers* with 2 or 3 layers of standard silicon hybrid planar pixel sensors at radii of ~ 8-35 cm and an *inner pixel layers* with 1-2 layers of very radiation hardness pixel sensors (Si-3D, Si-thin-planar and diamond are beginning investigated as valid option) at the radii below ~ 8 cm. For what concerns the intermediate and outer radii five silicon n-in-p micro-strip layers are foreseen: the innermost 3 layers will be made of *short-strips* (24 mm strip length) and the outer 2 layers of *long-strips* (95 mm strip length). In addition of that ATLAS planned to go through different upgrades for TDAQ, L1-Trigger and Calorimeter electronics.

## 2. ATLAS IBL

The ATLAS IBL will be the fourth layer added to the present Pixel Detector between a new beam pipe and the current inner Pixel Detector layer (*B*-layer). It consists of 14 tilted staves (64 cm long 2 cm wide and tilted in $\phi$ of 14°) equipped with 32 front-end chips per stave and sensors facing the beam pipe. The inner radius of IBL will be of 31 mm with the outer radius of 40 mm while the sensor will present a mean radius of 33 mm. The front-end chip foreseen for the IBL is called FE-I4 and it is described in [5]. The FE-I4 chip was designed in 130 nm CMOS technology and it consists of 26880 pixel cells organized in a matrix of 80 columns (on 50 μm pitch) by 336 rows (on 250 μm pitch). Each front-end cell contains an independent, free running amplification stage with adjustable shaping, followed by a discriminator with independently adjustable threshold. The FE-I4 keeps tracks of the firing time of each discriminator as the time over threshold (ToT) with 4-bit resolution, in counts of an external supplied clock of 40 MHz nominal. A common sensor baseline for engineering and system purpose was chosen for the pixel module considering that there are two different silicon sensor technologies: planar n-in-n [5] manufactured by CiS (Germany) and 3D with passing through columns [5] manufactured by FBK (Italy) and CNM (Spain). Two different format layouts were foreseen for the two sensor types and they are shown in Figure 1. The basic unit of the IBL is a module that consists of two (for planar, see Fig.1-a) or one (for 3D, see Fig.1-b) front-end chips bump bonded to one sensor. For single-chip (two-chip) assemblies the nominal active coverage for particles normal to the beam is 98.8% (97.4%). An air gap of 100 μm (200 μm) between single-chip (two-chip) assemblies has been assumed to take into account the higher bias voltages needed by two-chip planar sensor with respect to 3D assemblies.

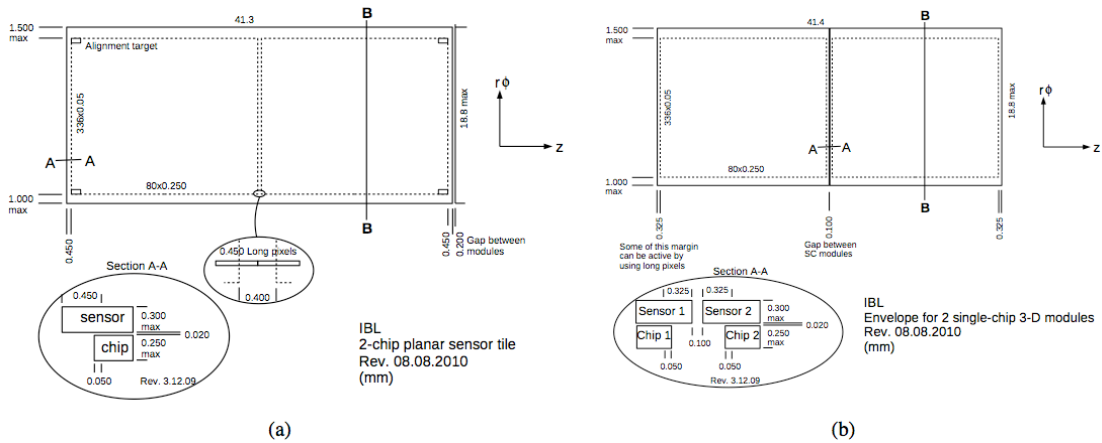

**Figure 1**: Module format layout for different sensor types: (a) two-chip module foreseen for planar sensors and (b) one-chip module foreseen for 3D sensors.

Due to the high radiation levels requirement foreseen for the IBL (NIEL dose: $5 \times 10^{15}$ 1MeV $n_{eq}/cm^2$ and TID: 250 Mrad [5]) irradiation and test-beam campaigns have been planed with single-chip assemblies for both sensor-type technologies.

## 2.1 Irradiation and beam tests qualification program

In order to study the radiation tolerance of the IBL modules a few irradiation campaigns were carried out. Several devices were irradiated in different facilities with different target fluencies. Two proton irradiation campaigns have been performed: one at Karlsruhe Institute of Technology (KIT) with a nominal proton beam energy of 25 MeV[1] up to $6 \times 10^{15}$ 1MeV $n_{eq}/cm^2$ and another at CERN PS with 24 GeV proton beam up to $5 \times 10^{15}$ 1MeV $n_{eq}/cm^2$. In addition other modules were irradiated with the TRIGA neutron reactor up to $5 \times 10^{15}$ 1MeV $n_{eq}/cm^2$ at Jozef Stefan Institute (JSI). In order to qualify the two sensor-type technologies three beam tests have been organized: two at DESY with 4 GeV electron beams and one at CERN SPS with 180 GeV pion beams. Few modules from both sensor-type technologies have been tested simultaneously in the beam in order to be able to assess their tracking and signal performance under identical condition with particular emphasis on modules that have been irradiated to the nominal IBL fluencies. In terms of operation both sensor technologies have been demonstrated to satisfy the IBL performance requirements and a detailed description of the test-beam measurement results are reported in [6][7]. Figure 2 shows the ToT distribution for a planar and a 3D sample irradiated up to $6 \times 10^{15}$ 1MeV $n_{eq}/cm^2$. Due to the different front-end tunings (planar (Fig.2-a): 5 ToT at 20 ke with a threshold of 1400e; 3D (Fig.2-b): 8 ToT at 20 ke with a threshold of 2950e) it makes impossible the direct comparison between the two distributions.

---

[1] Due to the low energies the ionizing radiation damage to the front-end went beyond the requirements (250 Mrad). Estimated Total Ionizing Dose about 750 Mrad for a fluence of $5 \times 10^{15}$ 1MeV $n_{eq}/cm^2$.

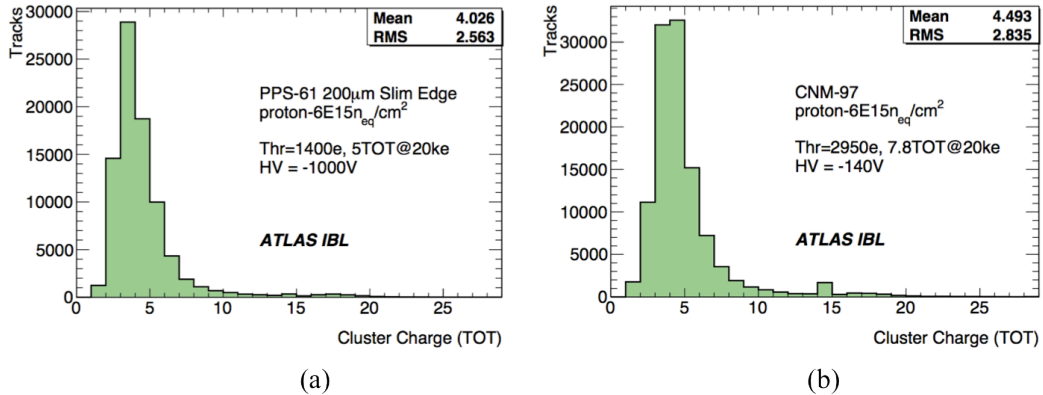

(a)     (b)

**Figure 2**: ToT distribution for (a) planar silicon sensor irradiated up to $6 \times 10^{15}$ 1MeV $n_{eq}/cm^2$ and biased at 1000V and (b) 3D silicon sensor irradiated up to $6 \times 10^{15}$ 1MeV $n_{eq}/cm^2$ and biased at 140V tested at CERN SPS with 180 GeV pion beams. Due to the different front-end tunings (planar: 5 ToT at 20 ke with a threshold of 1400e; 3D: 8 ToT at 20 ke with a threshold of 2950e) it makes impossible the direct comparison between the two distributions.

## 3. Summary

Following the LHC upgrade phases towards HL-LHC, ATLAS planned to upgrade the Inner Detector in two phases. The first will be in the 2013-14 with the installation of an additional innermost pixel layer so-called Insertable B-Layer (IBL) to maintain an excellent vertex detector performance and compensate possible inefficiencies of the current Pixel detector. To cope with the luminosity foreseen for HL-LHC a complete replacement of the Inner Detector is planned for the ~2022 shutdown with an all silicon system (pixel and micro-strip). In addition of that significant upgrades to the TDAQ, L1-Trigger and Calorimeter electronics are foreseen.